# Updated optical design and trade-off study for MOONS, the Multi-Object Optical and Near Infrared spectrometer for the VLT

E. Oliva*[a], S. Todd[b], M. Cirasuolo[b], H. Schnetler[b], D. Lunney[b], P. Rees[b], A. Bianco[c], E. Diolaiti[d], D. Ferruzzi[a], M. Fisher[e], I. Guinouard[f], M. Iuzzolino[a], I. Parry[e], X. Sun[e], A. Tozzi[a], F. Vitali[g]

[a] INAF-Arcetri Observatory, largo E. Fermi 5, I-50125 Firenze, Italy;
[b] STCF UK-ATC, Royal Observatory, Blackford Hill, Edinburgh, EH9 3HJ, UK;
[c] INAF-Brera Observatory, via Brera 28, I-20121, Milano, Italy;
[d] INAF-Bologna Observatory, via Ranzani 1, I-40127 Bologna, Italy;
[e] Institute of Astronomy, University of Cambridge, Madingley Road, Cambridge CB3 0HA, UK;
[f] GEPI, Observatoire de Paris-CNRS-Université Paris Diderot, 5 Pl. Janssen, 92195 Meudon, France;
[g] INAF-Rome Observatory, via di Frascati 33, I-00040, Monteporzio Catone, Rome, Italy;

**ABSTRACT**

This paper presents the latest optical design for the MOONS triple-arm spectrographs. MOONS will be a Multi-Object Optical and Near-infrared Spectrograph and will be installed on one of the European Southern Observatory (ESO) Very Large Telescopes (VLT). Included in this paper is a trade-off analysis of different types of collimators, cameras, dichroics and filters.

## 1. INTRODUCTION

MOONS is the acronym of Multi-Object Optical and Near-infrared Spectrograph which is currently being developed for the European Southern Observatory (ESO). The instrument was selected by ESO[1] following a call in 2010 for a wide field spectroscopic instrument concepts among the ESO community. In 2013, after the phase-A concept design review, the Science and Technology Committee (STC) of ESO recommended that the instrument should be developed, build and installed. The kick-off meeting to start the preliminary design phase is imminent. The main scientific aims[2] include galactic archeology (based on a detailed analysis of a few millions single stars in the Milky Way and in the Local Group) and cosmology (based on a spectroscopic study of the integrated light from several millions of galaxies at redshifts z>1). The top-level requirements for the instrument that directly affect the design of the spectrograph are:

- Simultaneous spectroscopy of 800 (goal 1000) objects/sky fibers at one of the VLT foci.
- Sky-projected diameter of each fiber 1.0 (goal 1.2) arc-sec.
- Medium spectral resolution mode (R>3,000, goal R> 4,000) with simultaneous spectral coverage over 0.8 - 1.8 µm.
- High spectral resolution mode covering parts of the J, H bands (at R>18,000; goal R=23,000) and the CaII triplet (at R>6,000; goal R=8,000).
- Cross-talk between neighboring spectra lower than 2% (goal 1%).

These requirements translate into remarkable challenges for the design of the spectrograph. A first analysis of the spectrometer parameters and optical design, used for the phase-A proposal, was already published.[3] Here we present an updated study and a trade-off analysis between different types of collimators, cameras, dichroics and filters. The new, provisional design fits into a significantly smaller volume, has a blue cut-off extended to lower wavelengths and a higher resolution mode that includes both the CaII and Oxygen triplets.

*oliva@arcetri.astro.it; phone +39 0552752291

## 2. OVERALL DESIGN OF THE SPECTROGRAPH

The fundamental scaling laws that determine the first-order parameters of the spectrographs were already described in a former paper[3]. The results are summarized in Table 1. The most striking and demanding aspects for the design of the spectrograph are
- Huge field of view for the collimator, equivalent to a 23' long-slit spectrometer on an 8m telescope.
- Extremely fast cameras (about F/1) with huge field of view, equivalent to a 23' x 23' imager on an 8m telescope.
- Good image quality over most of the detector: spots with $D_{rms}$<30 µm and >95% of encircled energy in D<75 µm.
- High resolution-slit product, achievable only with a large collimated beam that severely limits the use of lenses.
- Simultaneous availability of low and high spectral resolution modes, with simple and affordable exchange modes.

**Table 1** Main parameters of the MOONS spectrometers

| Parameter | Value | Comment |
|---|---|---|
| Number of spectrometers | 2 | Impossible to fit >800 fibers in a single spectrograph |
| Number of fibers per spectrometer | 512 | A dozen fibers may be permanently fed by calibration sources |
| Fiber diameter | 0.15 mm | Physical size |
| | 1.05" | Sky-projected angle |
| | 3 pixels | Projected size on detector |
| Distance between fibers | 0.40 mm | Physical size |
| | 5 pixels | Projected size on detector |
| Input slit to spectrometer | F/3.5 | Angular aperture of beam |
| | 200 mm | Physical length of the slit |
| | 23 arc-min | Equivalent slit length in sky-projected angles |
| Diameter of collimated beam | 265 mm | Set by requirements on resolving power, limited by cost/feasibility of VPH dispersers and by constraints on the overall size, volume and mass of the spectrometer. |
| Dispersers | VPH + prisms | See Section 5 |
| Number of arms/cameras | 3 | Three wavelength ranges, split via dichroics (see Section 6) |
| Cameras focal aperture | F/1.04 | Set by detector size, wavelength coverage and resolving power |
| Detectors | 4096 x 4096 | H4RG for the YJ and H arms |
| | 15µm pixel | CCD for the optical RI arm |

Compared to the phase-A analysis, we studied the possibility to extend the spectral coverage toward the blue and increase the resolving power around the CaII triplet, also including the region of the Oxygen triplet at 0.77µm, very important for the study of low-metallicity stars. A provisional layout including these extended capabilities is shown in Figure 1. It includes the smallest and more cost-effective optics; other types of collimators and cameras are discussed in Sections 3 to 4. The spectral coverage and resolutions are summarized in Table 2.

**Table 2** Spectral coverage and resolutions of a possible spectrometer with extended capabilities.

| | | |
|---|---|---|
| Spectral coverage and resolving power in the MR configurations | 0.70-0.95 µm | RI, R=4,600 at central wavelength |
| | 0.95-1.35 µm | YJ, R=4,200 at central wavelength |
| | 1.45-1.81 µm | H, R=6,300 at central wavelength |
| Spectral coverage and resolving power in the HR configurations | 0.765-0.895 µm | HR-I, R=9,200 at central wavelength |
| | 1.177-1.268 µm | HR-J, R=19,800 at central wavelength |
| | 1.521-1.635 µm | HR-H, R=19,700 at central wavelength |

The light from the fibers is collimated by a spherical mirror. The input fibers are organized along a curved slit as depicted in Figure 2. The curvature in the YZ plane matches the curved focal plane of the collimator, while the curvature in the XY plane is used to compensate the curvature of the slit image introduced by the dispersers. The curvature radius in XY is chosen to obtain a straight slit image at the center of the H spectrum at medium resolution. The residual curvature in the other MR and HR configurations is quite small, the maximum value being 80 pixels peak-to-peak.

The collimated beam is split into three arms by two dichroic beam-splitters. The longest wavelengths (λ>1.4 μm) are reflected by the first dichroic toward the H arm. The shortest wavelengths ((λ<0.95 μm) are reflected by the second dichroic toward the RI arm. The remaining light (0.95 μm <λ<1.40 μm) passes through the second dichroic and feeds the YJ arm. The H arm also includes an order-sorter filter (see Section 6).

Each arm includes two interchangeable dispersive systems, made by a combination of VPH gratings and prisms; the advantages of this solution are discussed in Section 5. The dispersed light is re-imaged onto the detector by a camera. The optical elements of the cameras shown in Figure 1 are identical in the three arms. This is a specific advantage of this type of reflective cameras (see Section 4).

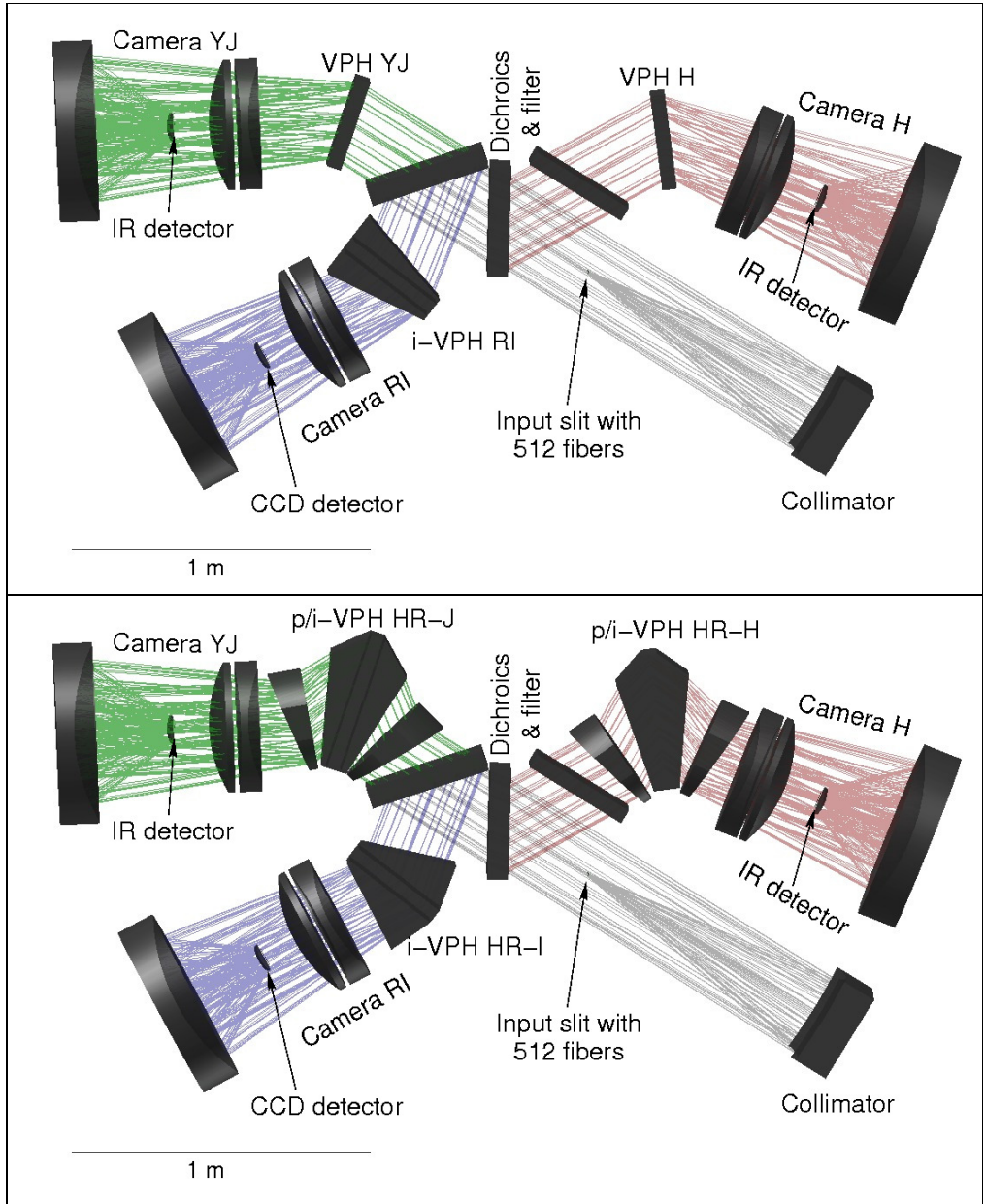

**Figure 1** Layout and rays-tracing of a possible spectrograph with extended capabilities. Top panel shows the setup in the medium resolution configuration, while the high resolution configuration is in the lower panel.

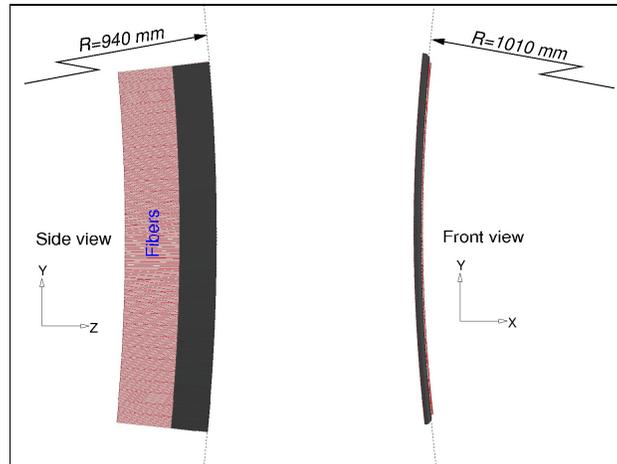

**Figure 2** Views of the input slit of the spectrometer. The fibers-axis are perpendicular to the cylindrical surface with R=940mm

## 3. TRADE-OFF ANALYSIS OF COLLIMATORS

Figure 3 shows the layouts of the types of collimators that were considered in the trade-off analysis. In the following we describe their characteristics. Their most important parameters are summarized in Table 3.

ON-AXIS. The collimator consists of a single spherical mirror working on-axis. The input slit is positioned at the focus of the mirror. The collimated beam is not optically corrected, but its aberrations do not influence the optical performances of the dispersers. The aberrations can be corrected in the cameras, without any significant increase of complexity, size and cost of the camera elements. Such compensation works only if the illumination foot-prints on the camera elements are similar in the medium and high resolution configurations. This has important consequences on the design of the dispersers (see Section 5). This type of collimator was used in the APOGEE[4] spectrometer. Its main advantages are simplicity, limited size and cost. The main drawback is the obscuration the collimated beam by the input slit. However, with a careful mechanical design, the fraction of vignetted beam can be limited to ~1.9%. Other drawbacks are the difficulty to insert mechanical elements at the slit (e.g. a shutter) and the increased complexity in the optical testing of the cameras.

RFL-SCHMIDT. The collimator consists of an off-axis section of a reflective Schmidt camera. The primary mirror is spherical. The secondary mirror is a flat that folds the beam to minimize the length of the system. The tertiary mirror is a canonical Schmidt aspheric corrector. The input slit does not interfere with the beam, i.e. there is no vignetting. The collimated beam is free from aberrations; this simplifies the design and optical testing of the cameras. The main drawback is the large mass and cost of the optics. The primary mirror is much higher than its on-axis equivalent because the pupil image is much farther from the input slit. Other solutions with three mirrors, including canonical TMA's, can be found by adding power to the secondary mirror. They all have similar sizes and volumes, because of the huge field of view and requirement on the position of the pupil. Their costs are similar or higher than the reflective Schmidt, because of the higher complexity of the surfaces to manufacture.

MAKSUTOV-CHROMATIC. The collimator consists of an off-axis section of a Maksutov camera. The primary mirror is spherical. The diverging corrector is a large singlet lens of fused-silica with spherical surfaces. Its overall size and cost are intermediate between those of the other two collimators. The collimated beam is corrected for all aberrations except chromatism, a canonical design with an achromatic doublet is practically impossible because of the limited availability of glasses of large enough size and transmitting in the required wavelengths range. The chromatic aberration can be corrected in the cameras. Such an approach, also known as "4C" (collimator compensation of camera chromatism[6]) was used in X-SHOOTER[5] to simplify the design of the cameras. However it does not help in our case, i.e. all the cameras designed with this collimator turned out to have similar performances, sizes, volumes and costs as those with the other collimators.

**Table 3** Parameters of the collimators considered in the design of the MOONS spectrometer

| Name | Length (m) | Height of largest element[1] (m) | Mass[2] (kg) | Relative cost[3] | Comments |
|---|---|---|---|---|---|
| ON-AXIS | 1.2 | 0.68 | 47 | 1.0 | Input slits obscures ~1.9% of the collimated beam |
| RFL-SCHMIDT | 1.1 | 1.00 | 175 | 4.7 | |
| MAKSUTOV | 1.3 | 0.84 | 105 | 3.7 | Diameter of corrector lens = 0.7 m |

*Notes to table*
(1) The height is measured perpendicular to the plane of Figure 3.
(2) Total mass of the optics in one collimator.
(3) Relative costs, ROM estimates.

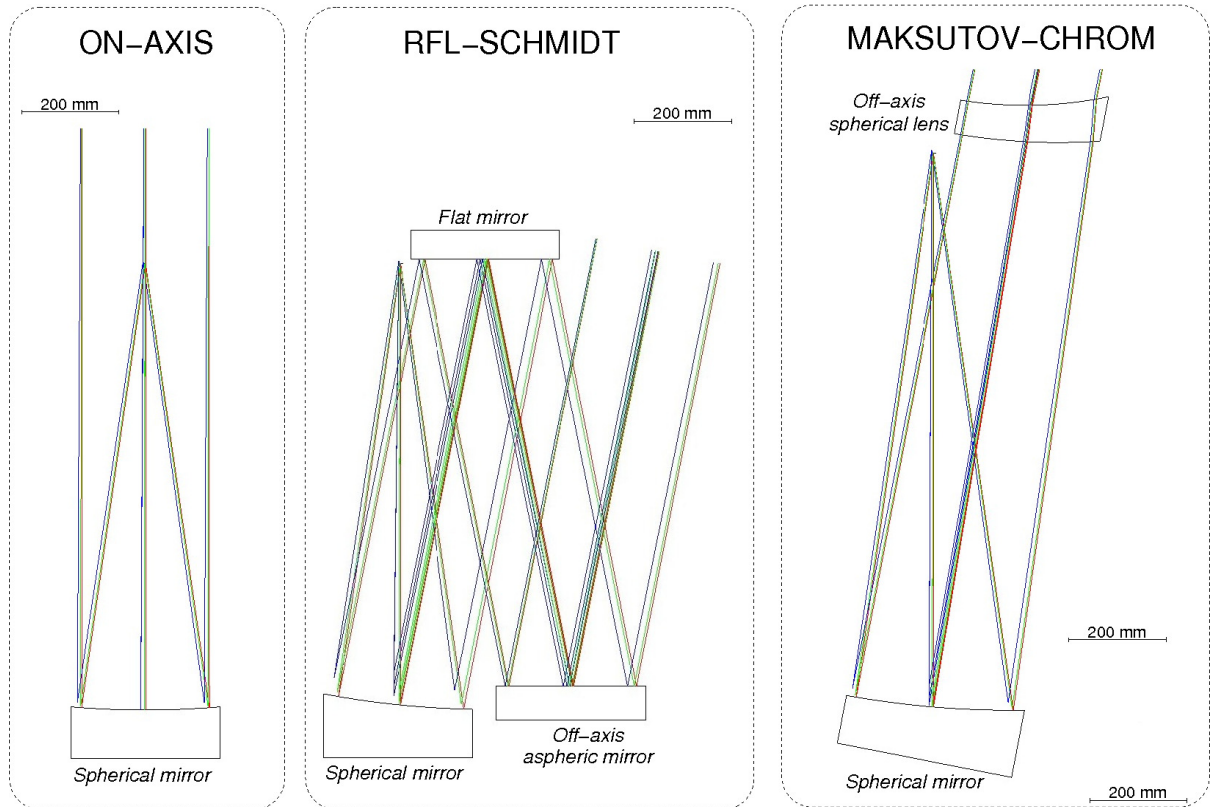

**Figure 3** Layouts of the different types of collimators considered for the design of the MOONS spectrometer.

## 4. TRADE-OFF ANALYSIS OF CAMERAS

The cameras are large, because of the large diameter of the collimated beam, and extremely fast: about F/1. Their design is particularly challenging because of the very limited choice of optical materials available in large sizes and transparent over the required wavelengths range. Table 4 summarizes the main parameters of the optical materials which are practically available. Of these, only three are transparent over the whole spectral range; they all have quite low refractive indices.

**Table 4** Selected properties of optical materials available in large sizes and volumes

| Optical material | $n^{(1)}$ | TH-97%$^{(2)}$ (mm) | | | Relative cost$^{(3)}$ | Comment |
| --- | --- | --- | --- | --- | --- | --- |
| | | I (0.70-0.95) | YJ (0.95-1.35) | H (1.47-1.81) | | |
| BK7 glass | 1.51 | >300 | 80 | 15 | 1 | |
| Silicon mono-crystal | 3.52 | <1 | <1 | >300 | 2 | Max. diameter ~ 400 mm |
| F2 glass | 1.60 | >300 | 140 | 20 | 3 | |
| Fused silica IR-grade | 1.45 | >300 | >300 | >300 | 5 | |
| S-FPL51 glass | 1.49 | >300 | >300 | >300 | 6 | |
| $CaF_2$ | 1.43 | >300 | >300 | >300 | 12 | Limits on size and thickness |
| SF6 glass | 1.77 | >300 | 180 | 25 | 15 | |
| ZnSe | 2.47 | 40 | 80 | 190 | 30 | Max. blank thickness ~ 60mm |

*Notes to table*
(1) Refractive index at 1.2 microns.
(2) Thickness above which the minimum internal transmission within a given band drops below 97%.
(3) Relative cost per unit-volume. Values are indicative for blanks of large size

In the previous study[3] we concentrated on refractive cameras (only lenses). That design consisted of two quasi-identical cameras with 7 spherical lenses of low-cost glasses (6 lenses of BK7, 1 lens of F2 and 1 small lens of SF6) for the YJ and I arms, and one camera with 6 spherical lenses (1 fused silica, 5 of mono-crystalline Silicon) for the H arm. In all cameras the detector was tilted in the spectral direction, a strategy that largely simplifies the design[7]. The main drawbacks of these cameras are the internal absorption of the glasses in the J band and the relatively large number of optical surfaces. We therefore developed other designs with fewer lenses (including aspheric surfaces) and using the few materials transparent in the J band. The preliminary results yielded very expensive cameras, which are about 3 times those of the reflective design discussed in the following sections.

Here we concentrate on alternative, more cost-effective designs with reflective cameras (RFL). The layouts are visualized in Figure 4 and their main parameters are summarized in Table 5. Three lenses (two correctors and one field-flattener) are needed to achieve the required image quality. Given the low power of the lenses, it is relatively easy to design quasi-identical cameras (i.e. same optical elements with different distances) for all the arms. It is also possible to design cameras with all the lenses made of fused-silica. These cameras have the outstanding property of being temperature independent, i.e. they have the same optical performances at room and at cryogenic temperatures. All these facts significantly decrease the overall costs, and simplify the integration and tests of the cameras.

The only drawback of reflective cameras is the obscuration by the last lens and the detector that are positioned on-axis, close to the focus of the mirror. Given the fast aperture of the camera, the obscuration cannot be eliminated using off-axis systems (this would require designing a camera faster than F/0.4). The only way to mitigate the obscuration is using the two corrector lenses as beam-expander, to increase the size of the beam between the second lens and the mirror. This strategy is applied, at different levels, in the three RFL cameras shown in Figure 4. The diameter of the beams around the detector (290mm, 340mm and 380mm) is added as suffix to the names. Since the area of central obscuration is the same for all cameras, the fraction of vignetted light decreases with the second power of the beam diameter.

All RFL cameras require 3 aspheric surfaces: L1, L3 and the mirror. The design becomes easier (i.e. milder aspherics and possibility to include flat surfaces) going to larger beam sizes. The main drawbacks of the larger cameras are size, mass and cost.

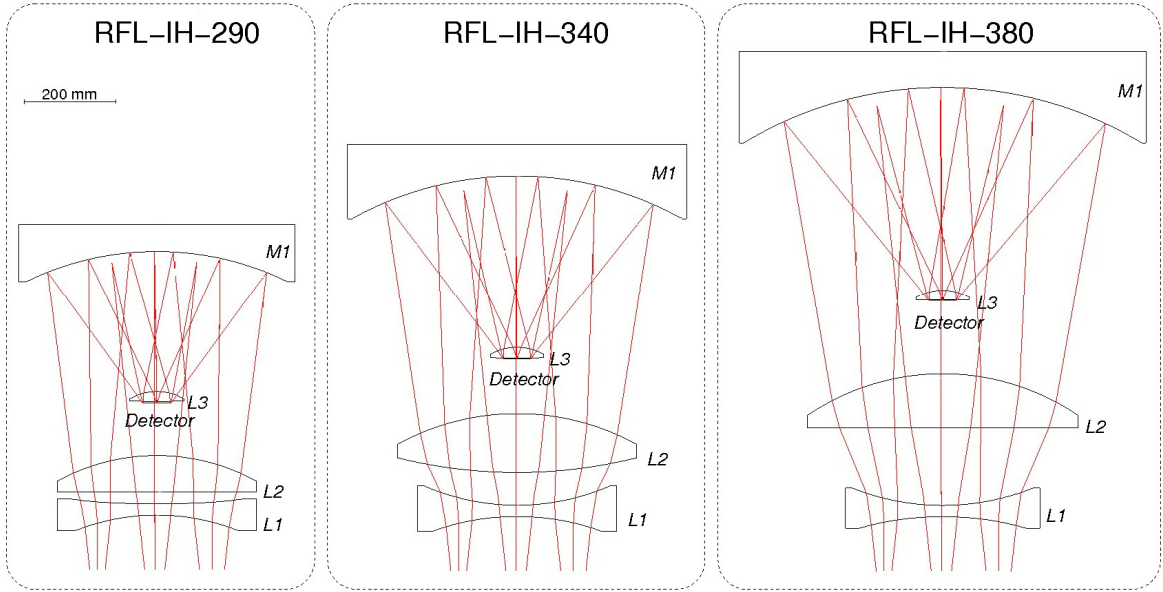

**Figure 4** Layouts of different types of reflective cameras. Light-path is from bottom to top, i.e. L1 - L2 - M1 - L3.

**Table 5** Parameters of the reflective cameras considered in the design of the MOONS spectrometer

| Name | Length (mm) | Diameter of largest lens (mm) | Diameter of mirror (mm) | Mass[1] (kg) | Relative cost[2] | Central Obs.[3] (%) | Comments |
|---|---|---|---|---|---|---|---|
| RFL-IH-290 | 650 | 440 | 610 | 110 | 1.0 | 17 | Same camera for all arms |
| RFL-IH-340 | 830 | 520 | 740 | 200 | 1.5 | 13 | Same camera for all arms |
| RFL-IH-380 | 1030 | 620 | 920 | 300 | 1.9 | 10 | Same camera for all arms |

*Notes to table*
(1) Total mass of the optics in one camera.
(2) Relative costs, ROM estimates. The mirrors are made of low-CTE glass whose cost is similar to BK7.
(3) Fraction of light vignetted by L3+detector.

## 5. DISPERSERS

Transmission VPH-gratings are the ideal (and indeed the only) type of dispersers that can handle the huge field of view of the spectrometer, maintaining a high throughput. To avoid spectral ghosts produced by multi-reflections in VPHs[8], the gratings work with an off-axis angle of 4 degrees, relative to the Littrow configuration. To optimize the optical performances and minimize the size of the disperser, the gratings are positioned close to the pupil image; this requirement becomes imperative with the on-axis collimator, because the camera must compensate the aberrations introduced by the collimator. The on-axis collimator also requires that the MR and HR gratings must have similar anamorphic magnification, otherwise the camera cannot compensate the aberrations introduced by the collimator. This constraint rules out designs where the MR/HR exchange mechanism is made by changing the angle between the camera and the collimated beams. Constant-anamorphism also implies that the curvature of the slit image on the detector does not significantly change in the MR and HR configurations.

Given these constraints, the most convenient approach to change the resolving power is using a combination of prisms and VPH-gratings. The prisms are used to change the incidence/diffracted angles of the chief-ray on the grating, while maintaining the same input angle (from the collimator) and output angle (to the camera). The prisms parameters can also be tuned to obtain similar anamorphic magnifications in the MR and HR configurations. Moreover, the prisms are useful to decrease the size/length of the gratings, i.e. they can simplify the manufacturing and decrease the costs of the HR gratings. Finally, the change of resolving power with the prism-VPH combinations does not require highly accurate mechanical positioning mechanisms (see Figure 5).

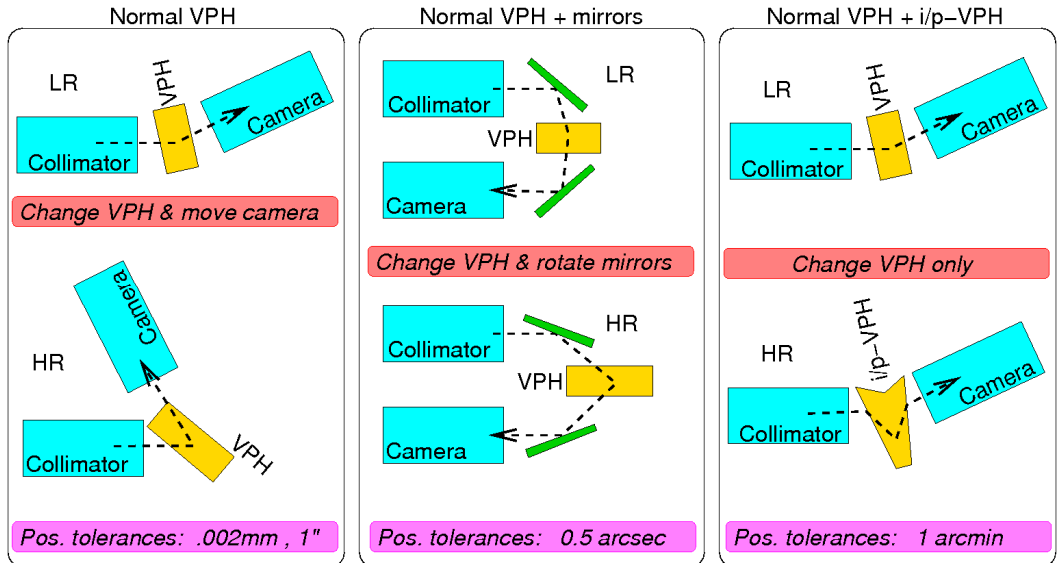

**Figure 5** Positioning tolerances to achieve a 0.1 pixel repeatability when changing resolving power configurations.

The following combinations of prism-gratings can be used.
- "i-VPH", consisting of two prisms directly glued onto the VPH grating. To optimize throughput and avoid thermal-induced mechanical stresses, the prisms must be of the same material as the grating.
- "p-VPH", consisting of a VPH grating in between two separate prisms. In this configuration one can use prisms with high refractive index to achieve high resolving powers even starting from a geometrical configuration with small angles between the collimator and the camera.
- "i/p-VPH", consisting of a i-VPH immersed grating in between two separate prisms.

In the RI arm, the change of resolving between MR and HR configurations is a factor of two (see Table 2). This can be achieved using two i-VPH's with opposite angles directions (see Figure 1).

In the H and YJ arms the ratio between the resolving powers in the HR and MR configurations are much larger, i.e. a factor of 3.2 and 4.7 (see Table 2). Such dynamic ranges can only be achieved using p-VPHs with prisms of mono-crystalline Silicon whose very high refractive index (n~3.5) allows one to achieve large deviations with relatively thin prisms. Other optical materials would require much thicker prisms which are either unfeasible for practical limitations on the material (e.g. ZnSe) or too bulky to fit into the available space. Two types of prisms-grating combinations can be used. The first uses i/p-VPHs (left panels of Figure 6), while the second uses p-VPHs (right-hand panels of Figure 6). The p/i-VPHs are more massive and expensive, but have the advantage of achieving a resolving power ~10% higher than p-VPHs, for a given length of the grating. They also have lower incidence angles of the rays onto the gratings, i.e. lower reflection losses.

The transparency of the Silicon prisms in the HR-J configuration is a potentially critical issue. The internal transmission of Silicon at these short wavelengths depends strongly on the temperature. At room temperature the opacity is very high. An extrapolation of the available data indicate that the absorption should become negligible at T<100 K. However, direct measurements are missing.

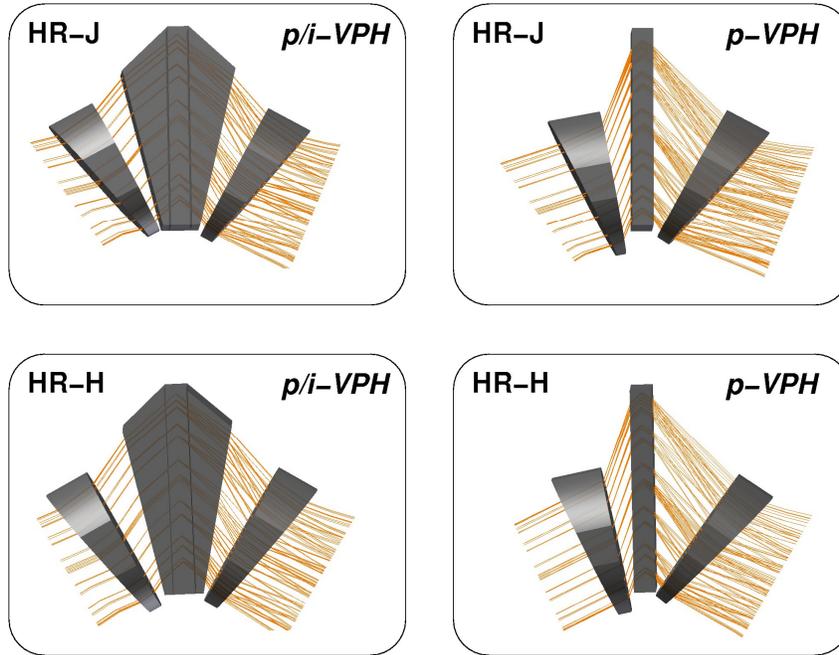

**Figure 6** Layouts and rays-tracing of different combinations of prisms and VPH gratings for the high resolution configurations. Light goes from left to right.

## 6. DICHROICS AND FILTERS

The cut-on/off wavelengths of the dichroic beam-splitters were specifically selected to coincide with regions of bad telluric transmission (see Figure 7). Such an approach optimizes the scientific output and simplifies the design, because it relaxes the requirements on the cut-on/off slopes. Luckily, the positions of the strongest atmospheric absorption bands are compatible with the top level requirements on the resolving powers in the three arms.

The design in Figure 1 is not ideal for the manufacturing of the first beam-splitter, because it transmits short and reflects longer wavelengths. Normally, dichroic coatings are much easier to fabricate if the short wavelengths are reflected, and the longer are transmitted. However, changing the arms-layout to simplify the dichroics, i.e. from H-RI-YJ to RI-YJ-H, makes it impossible to fit the HR-H and HR-J dispersers into the available space around the pupil images. In other words, the large HR-H and HR-J dispersers can only be accommodated in the first and last arms of the spectrograph.

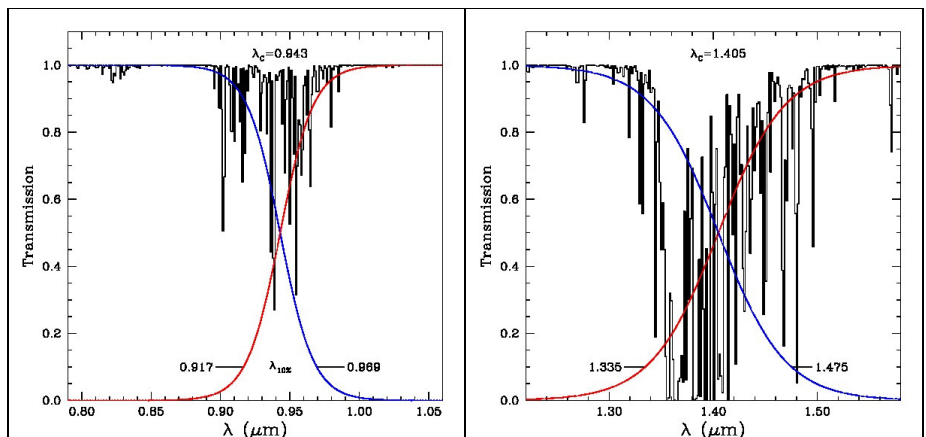

**Figure 7** Telluric transmission curves (black lines) and proposed transmission curves for the MOONS dichroics.

The grating dispersers work in first order. Order-sorter, (i.e. long-pass) filters are needed to block the short wavelengths that would otherwise reach the detector in second and higher orders, and contaminate the spectrum.
The first order-sorter filter is positioned at the entrance of the H-arm (see Figure 1). It could simply consist of a disk of mono-crystalline Silicon with A/R coating, thus achieving very high throughput (>99%) at low costs.
The order-sorter filter for the YJ and RI arms could be, most conveniently, integrated in the optics that feed the fibers. The simplest and most effective solution is fabricating the micro-lenses[9] with a commercial colored-glass (e.g. RG695).

Blocking of thermal radiation at 2.0 μm < λ < 2.6 μm could also be an issue, if the IR detectors are sensitive up to 2.5 microns. The fibers are relatively opaque in this wavelengths range and, therefore, provide some filtering. The amount of absorption depends on the length of fibers at cold temperatures; the results are displayed in Figure 8. The remaining thermal radiation can be filtered by imposing that the two dichroics have high reflectivity in this wavelengths range, and adding a cut-off (short-pass) coating to the H order-sorter filter.

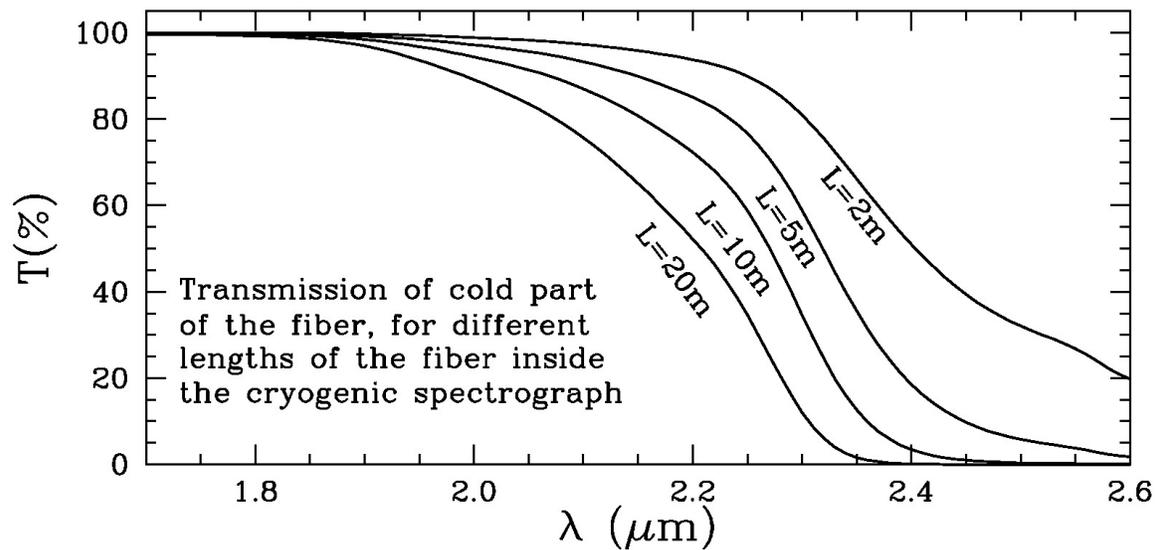

**Figure 8** Internal transmission of telecom fibers in the range of interest for blocking of the thermal radiation.

## 7. ACKNOWLEDGEMENTS

This work was financially supported by the Italian Institute of Astrophysics through the grants "TECNO-INAF-2011" and "VLT-PREMIALE-2011"

## REFERENCES

[1] Ramsay, S.; Hammersley, P.; Pasquini, L.; The Messenger, 145, 10 (2011)
[2] Cirasuolo, M.; Afonso, J.; Bender, R.; Bonifacio, P.; Evans, C.; Kaper, L.; Oliva, E.; Vanzi, L.; The Messenger, 145, 11 (2011)
[3] Oliva, E.; Diolaiti, E.; Garilli, B.; et al.; SPIE, 8445-187 (2012)
[4] Wilson, J. C.; Hearty, F.; Skrutskie, F.; et al.; SPIE, 7735-46 (2010)
[5] Vernet, J.; Dekker, H.; D'Odorico, S.; et al.; Astronomy & Astrophysics, 536, 105 (2011)
[6] Delabre, B.; Dekker, H.; D'Odorico, S.; Merkle, F.; SPIE, 1055, 340 (1989)
[7] Delabre, B.; Manescau, A.; SPIE, 7735-182 (2010)
[8] Burgh, E.B.; Bershady, M. A.; Westfall, K.B.; Nordsieck, H.; PASP, 119, 1069 (2007)
[9] Guinouard, I.; Amans, J.P.; Cirasuolo, M.; et al.; SPIE, 9151-180 (2014)